\documentclass[11pt]{article}

\usepackage{amsmath,amssymb}
\usepackage{graphicx,color}
\usepackage{mathrsfs}

\newcommand{\fundl}
{\setlength{\unitlength}{0.6pt}
\begin{picture}(10,10)
\put(0,0){\line(1,0){10}}
\put(0,10){\line(1,0){10}}
\put(0,0){\line(0,1){10}}
\put(10,0){\line(0,1){10}}
\end{picture}}

\newcommand{\sfund}
{\setlength{\unitlength}{0.3pt}
\begin{picture}(10,10)
\put(0,0){\line(1,0){10}}
\put(0,10){\line(1,0){10}}
\put(0,0){\line(0,1){10}}
\put(10,0){\line(0,1){10}}
\end{picture}}

\newcommand{\anti}
{\setlength{\unitlength}{0.55pt}
{\begin{picture}(16,22)
\put(4,-6){\line(0,1){20}}
\put(4,14){\line(1,0){10}}
\put(4,4){\line(1,0){10}}
\put(4,-6){\line(1,0){10}}
\put(14,-6){\line(0,1){20}}
\end{picture}}}

\newcommand{\symm}
{\setlength{\unitlength}{0.55pt}
{\begin{picture}(24,14)
\put(0,0){\line(0,1){10}}
\put(0,10){\line(1,0){20}}
\put(0,0){\line(1,0){20}}
\put(10,0){\line(0,1){10}}
\put(20,0){\line(0,1){10}}
\end{picture}}}

\newcommand{\santi}
{\setlength{\unitlength}{0.3pt}
{\begin{picture}(16,22)
\put(4,-6){\line(0,1){20}}
\put(4,14){\line(1,0){10}}
\put(4,4){\line(1,0){10}}
\put(4,-6){\line(1,0){10}}
\put(14,-6){\line(0,1){20}}
\end{picture}}}

\newcommand{\ssymm}
{\setlength{\unitlength}{0.3pt}
{\begin{picture}(24,14)
\put(0,0){\line(0,1){10}}
\put(0,10){\line(1,0){20}}
\put(0,0){\line(1,0){20}}
\put(10,0){\line(0,1){10}}
\put(20,0){\line(0,1){10}}
\end{picture}}}

\numberwithin{equation}{section}
\begin{document}

\quad 
\vspace{-2.0cm}

\begin{flushright}
\parbox{3cm}
{
{\bf Dec. 2009}\hfill \\
YITP-09-111 \hfill \\
 }
\end{flushright}

\vspace*{2.5cm}

\begin{center}
\LARGE\bf 
{\fontfamily{pbk}{\bfseries{\scshape{\selectfont 
On AGT Conjecture for Pure Super Yang-Mills \\and W-algebra}}}}\\
\end{center}
\vspace*{1.0cm}
\centerline{\large 
\textbf{Masato Taki}
}
\begin{center}
{\textit{Yukawa Institute for Theoretical Physics, Kyoto University, Kyoto 606-8502, Japan}}
%

\begin{center}
{taki@yukawa.kyoto-u.ac.jp}\\
\end{center}

\end{center}

\vspace*{0.7cm}

\centerline{\bf \textbf{Abstract}} 

\vspace*{0.5cm}
Recently Alday, Gaiotto and Tachikawa have proposed relation between 2- and 4-dimensional conformal field theories.
The relation implies that 
the Nekrasov partition functions of $\mathcal{N}=2$ superconformal gauge theories
are equal to conformal blocks associated with the conformal algebra.
Likewise,  a counterpart in pure super Yang-Mills theory exists in conformal field theory.
We propose a simple relation between the Shapovalov matrix of the $\mathcal{W}_3$-algebra and 
the Nekrasov partition function of $\mathcal{N}=2$ $SU(3)$ Yang-Mills theory.
\vspace*{1.0cm}

\vfill

\thispagestyle{empty}
\setcounter{page}{0}

\newpage

\section{Introduction}
The study of conformal field theory (CFT) in 2-dimensions \cite{Belavin:1984vu} 
has been an important subject in physics and mathematics,
since it is available for various applications such as string theory, the critical phenomena and the representation theory of groups.

$\mathcal{N}=2$ gauge theories are also important topics of quantum field theory.
It is known that these systems are exactly solvable by using the Seiberg-Witten theory.
The microscopic verification of the theory had been given by Nekrasov \cite{Nekrasov:2002qd}\cite{Nekrasov:2003rj}.
He showed that the integral over the ADHM moduli space of instantons implies the generating function
of the Seiberg-Witten prepotential, which is called the Nekrasov partition function.
The localization fomula reduces the integral over the instanton moduli space to the instanton counting problem.
The resulting Nekrasov partition function has a combinatorial expression,
which is summation over Young diagrams.

Recently, remarkable relation 
between 2- and 4-dimensonal CFT's was proposed by Alday, Gaiotto and Tachikawa \cite{Alday:2009aq}.
The core of their proposal is that the instanton part of the Nekrasov partition for a $SU(2)$ superconformal gauge theory
is exactly equal to a certain conformal block of 2-dimansional CFT.
This means that we can recast the Nekrasov partition function into a CFT correlator by summing over Young diagrams.

In this paper, we study relation between $\mathcal{N}=2$  $SU(3)$ pure Yang-Mills theory and conformal symmetry.
In \cite{Gaiotto:2009ma} Gaiotto proposed that the Nekrasov partition function of $SU(2)$ pure Yang-Mills theory
can be represented as the norm of a certain state in the Virasoro Verma module.
Moreover Mironov and Morozov showed that the Shapovalov form of the Verma module gives this state by applying the
decoupling limit of flavors to the Alday-Gaiotto-Tachikawa conjecture (AGT conjecture) \cite{Marshakov:2009gn}.
We study the decoupling limit of the Wyllard proposal \cite{Wyllard:2009hg} which implies that a correlator of
$A_{N-1}$ Toda field theory with $\mathcal{W}_N$-symmetry \cite{Fateev:1987vh}\cite{Fateev:1987zh}
is related to $\mathcal{N}=2$  $SU(N)$ conformal quiver gauge theory.
We call it the AGT-W conjecture.
In \cite{Dijkgraaf:2009pc}\cite{Bonelli:2009zp}\cite{Alday:2009qq} 
string theory derivations were proposed.
We focus on the case of $\mathcal{W}_3$-symmetry \cite{Zamolodchikov:1985wn}.
In  \cite{Mironov:2009dr}\cite{Mironov:2009qt}\cite{Mironov:2009by} it was shown that the AGT-W conjecture implies that
 the 4-point conformal block associated with $\mathcal{W}_3$-symmetry is equal to the Nekrasov
 partition function of $SU(3)$ superconformal gauge theory.
We study the decoupling limit of $SU(3)$ superconformal gauge theory 
and the conformal block for $\mathcal{W}_3$-symmetry, and then we relate the Nekrasov partition function
 $SU(3)$ pure Yang-Mills to the Shapovalov form for the $\mathcal{W}_3$-algebra.
 The result implies that the Nekrasov partition functions may be deeply linked to the representation theory
 of conformal algebra.

This paper is organized as follows. In section 2, we give a brief reviews on the AGT conjecture.
In section 3, we propose a relation between  $SU(3)$ pure Yang-Mills and 
the Shapovalov matrix for the $\mathcal{W}_3$-algebra.
We check the proposal for 1-instanton partition function and the level-1 Shapovalov matrix.
By applying the decoupling limit of flavors, we show that the AGT-W conjecture implies our proposal.
Conclusions are found in section 4.
 In Appendix A, we summarize the definition and properties of the conformal blocks. 
 Our proposal is verified explicitly at 2-instanton level in Appendix.B.
\section{Non-conformal AGT Conjecture and Virasoro Algebra}
The main issue we study in this paper is the AGT relation 
between a Nekrasov partition function and a conformal blocks of 2-dimensional CFT.
As a guide to the latter part, we give a brief review on the AGT conjecture.
\subsection{Alday-Gaiotto-Tachikawa conjecture}
The Nekrasov partition function \cite{Nekrasov:2002qd} is a generating function of the Seiberg-Witten prepotential 
$Z^{\textrm{inst}}=\exp(\frac{1}{\epsilon_1\epsilon_2}
\mathcal{F}^{\textrm{SW}}+\cdots)$.
The explicit form of the partition function is found in the next section.
The parameters $\epsilon_{1,2}$ correspond to the so-called $\Omega$-background \cite{Nekrasov:2003rj}.

The Nekrasov partition function for the  $\mathcal{N}=2$ $SU(2)$ gauge theory with $4$-flavors
is a function of $8$ parameters:
\begin{align}
Z^{\textrm{inst}}(a,\vec{\mu} ,x,\epsilon_1,\epsilon_2)
=\sum_{k=0}^{\infty}x^{k}Z_{k}(a,\vec{m}, \epsilon_1,\epsilon_2).
\end{align}
While the $7$ parameters $a$, ${\mu}_f$, $\epsilon_1$ and $\epsilon_2$ have mass dimension one,
the factors $Z_k(a,\vec{m}, \epsilon_1,\epsilon_2)$ are dimensionless, 
which is a reflection of the fact that the gauge theory is conformal.
Let us introduce the scale $\hbar=\sqrt{-\epsilon_1\epsilon_2}$.
Then we can scale the patririon function as follows:
\begin{align}
Z^{\textrm{inst}}(a,\vec{\mu} ,x,\epsilon_1,\epsilon_2)
=\sum_{k=0}^{\infty}x^{k}Z_{k}\left(\frac{a}{\hbar},\vec{\mu}, e_1,e_2\right).
\end{align}
Here we introduce the dimensionless $\Omega$-background and masses:
\begin{align}
&e_{{}_{E}}=\frac{\epsilon_{{}_E}}{\hbar},\quad E=1,2\\
&\mu_f=\frac{m_f}{\hbar},\quad f=1,\cdots,4.
\end{align}

Let us introduce the following parametrization \cite{Alday:2009aq} of the gauge theory parameters:
\begin{align}
&c=1-6e^2\nonumber\\
&\Delta=\frac{a^2}{\hbar^2}-\frac{e^2}{4}
\end{align}
Here $e$ denotes $e=e_1+e_2$.
These new parameters play a role in the AGT relation.

A simplest version of the AGT conjecture implies that the above Nekrasov partition function 
coincides with the 4-point spherical conformal block
of the Virasoro algebra with central charge $c$ \cite{Marshakov:2009gs}:
\begin{align}
Z^{\textrm{inst}}\left(\frac{a}{\hbar},\vec{\mu},x, e_1,e_2\right)
=\mathcal{B}_{\Delta}
\tiny{\left[ {\begin{array}{cc}\Delta_1 \hspace{-0.2cm}& \Delta_2 \\ \Delta_3 \hspace{-0.2cm}& \Delta_4 \end{array}}  \right]}(x)
\end{align}
Here $\Delta_f=\alpha_f(e-\alpha_f)$ is the conformal dimension of the external states.
The external momentum $\alpha_f$ are corresponding to the mass parameters of the gauge theory:
\begin{align}
&\mu_1=\alpha_1+\alpha_2-\frac{e}{2} \nonumber\\
&\mu_2=-\alpha_1+\alpha_2+\frac{e}{2} \nonumber\\
&\mu_3=\alpha_3+\alpha_4-\frac{e}{2} \nonumber\\
&\mu_4=\alpha_3-\alpha_4+\frac{e}{2} \nonumber
\end{align}
See Appendix.\ref{sec:cb} for the definition of the conformal block $\mathcal{B}$.

In this way, the instanton counting of the $N_c=2$ $N_f=4$ gauge theory 
keeps the representation theory of the Virasoro algebra \cite{Marshakov:2009gs} behind.
It is very natural to expect that the Nekrasov partition functions of a$\mathcal{N}=2$
gauge theory and the representation theory of the symmetry algebra of 2-dimensional CFT are closely related
whether the gauge theory is conformal or not.
In the following we review Gaiotto's proposal on the $SU(2)$ pure supre Yang-Mills theory.

\subsection{Gaiotto conjecture (non-conformal AGT conjecture)}

Let us consider the Nekrasov partition function of the $SU(2)$ pure supre Yang-Mills theory:
\begin{align}
Z^{\textrm{inst}}(a ,\Lambda,\epsilon_1,\epsilon_2)
&=\sum_{k=0}^{\infty}\Lambda^{4k}Z_{k}(a,\epsilon_1,\epsilon_2)\nonumber\\
&=\sum_{k=0}^{\infty}\frac{\Lambda^{4k}}{(-\epsilon_1\epsilon_2)^{2k}} Z_{k}\left(\frac{a}{\hbar},e_1,e_2\right)
\end{align}
Notice that the factors $Z_k$ here have mass dimension $-4k$ in order that the full partition
function $Z^{\textrm{inst}}=1+\mathcal{O}(\Lambda^{4})$ is dimensionless.

In \cite{Gaiotto:2009ma} Gaiotto proposed that the partition function is realized as the norm of a certain state of the Virasoro algebra:
\begin{align}
Z^{\textrm{inst}}({a}{\hbar},\,\Lambda\hbar,\,e_1\hbar,\,e_2\hbar)=\langle \Delta, \Lambda |\Delta, \Lambda \rangle.
\end{align}
Here we eliminate the overall scale scale $\hbar$ from the parameters, 
and the conformal dimension of the internal state is $\Delta=a^2-e^2/4$. 
He found that the state $ |\Delta, \Lambda \rangle= |\Delta \rangle +\cdots$ must satisfy the following conditions:
\begin{align}
& |\Delta, \Lambda \rangle =\sum_{n=0} \Lambda^{2n}|\Delta,n \rangle \nonumber\\
& L_1  |\Delta,n \rangle = |\Delta,n-1 \rangle,\\
& L_k  |\Delta,n \rangle =0 \textrm{ for } k=2,3,4,\cdots,\nonumber
\end{align}
where $|\Delta\rangle$ is the primary state with conformal dimension $\Delta$.
It is not obvious whether such a state exists or not.
Marshakov and Mironov showed that the Gaiotto state $ |\Delta, \Lambda \rangle$ 
for pure $SU(2)$ super Yang-Mills theory 
is given by the Shapovalov matrix $Q_{\Delta}(Y;Y^{\prime})$ \cite{Marshakov:2009gn}\cite{Alba:2009fp}:
\begin{align}
 |\Delta,n \rangle=\sum_{|Y|=n} Q_{\Delta}^{-1}([1^n]; Y)\,L_{-Y}\,|\Delta \rangle .
\end{align}
Here $Y=\{Y_1,\, Y_2,\, \cdots\}=\left[1^{m_1}2^{m_2}\cdots \right]$ is a Young diagram with $|Y|=\sum Y_i=\sum j\,m_j$ boxes,
and $L_{-Y}$ denotes $L_{-Y_l}\cdots L_{-Y_2}\cdot L_{-Y_1}$.
The Shapovalov matrix is the following type of Gram matrix:
\begin{align}
Q_{\Delta}(Y;Y^{\prime})=\langle \Delta |L_{Y}L_{-Y^{\prime}} |\Delta \rangle.
\end{align}

The Gaiotto-Marshakov-Mironov proposal implies that the Nekrasov partition function for pure $SU(2)$ Yang-Mills is given by
\begin{align}
Z^{\textrm{inst}}({a}{\hbar},\,\Lambda\hbar,\,e_1\hbar,\,e_2\hbar)=\sum_{n} \Lambda^{4n}\, Q_{\Delta}^{-1}([1^n]; [1^n]).
\end{align}
Our interest in this paper has centered on this property of the instanton counting.
An important point is that the Nekrasov partition function of $SU(2)$ pure Yang-Mills
corresponds to a basic quantity in the representation theory of the Virasoro algebra.
Now a question arises; what is the counterpart in the instanton counting of $SU(N)$ pure super Yang-Mills theory?
In the next section we give an explicit answer for the question in the case of $SU(3)$ gauge theory.

\section{$SU(3)$ Pure Super Yang-Mills and $\mathcal{W}_3$-algebra}
In this section we propose a relation between the Nekrasov partition function of $SU(3)$ pure super Yang-Mills theory
and the Shapovalov matrix of the $\mathcal{W}_3$-algebra.
This proposal gives a nontrivial extension of the Gaiotto conjecture we reviewed in the previous section.
\subsection{Nekrasov formula}
The Nekrasov partition function is a generating function of the Seiberg-Witten prepotential.
It is given by the localization calculation of the path integral over the instanton moduli space with an appropriate measure.
The explicit form of the partition function for  $SU(N)$ pure super Yang-Mills theory is 
\cite{Flume:2002az}\cite{Bruzzo:2002xf}\cite{Nakajima:2003}
\begin{align}
\label{Nekrasov}
Z^{\textrm{inst}}(\vec{a},\Lambda,\epsilon_1,\epsilon_2)=\sum_{\vec{Y}}\frac{\Lambda^{2N_c |\vec{Y}|}}
{\prod_{\alpha,\beta=1}^{N_c}n_{\alpha,\beta}^{\vec{Y}}(\vec{a},\epsilon_1,\epsilon_2)}.
\end{align}
$\vec{Y}=(Y_1,\cdots,\,Y_N)$ is a vector consists of $N$ Young diagrams and its norm is defined by $|\vec {Y}|=\sum_n |Y_n|$.
Here the denominator is the eigenvalues of the torus action on the tangent space of the moduli space 
\cite{Flume:2002az}\cite{Bruzzo:2002xf}\cite{Nakajima:2003},
and it is given by the characteristic of the ADHM complex.
\begin{align}
\label{denominator}
n_{\alpha,\beta}^{\vec{Y}}(\vec{a},\epsilon_1,\epsilon_2)=&\prod_{(i,j)\in Y_{\alpha}} 
(-l_{Y_{\beta}}(i,j)\epsilon_1+(a_{Y_{\alpha}}(i,j)+1)\epsilon_2+a_{\alpha}-a_{\beta} )\nonumber\\
&\qquad\times 
\prod_{(i,j)\in Y_{\beta}} 
((l_{Y_{\alpha}}(i,j)+1)\epsilon_1-a_{Y_{\beta}}(i,j)\epsilon_2+a_{\alpha}-a_{\beta} ).
\end{align}
$\vec{a}=(a_1,\cdots,\, a_N)$ is the eigenvalue vector of the adjoint chiral fields.
An arm length and leg length of a Young diagram are defined by $a_Y(i,j)=Y_i-j$ and $l_Y(i,j)={Y^t}_j-i$.

We expand the partition function with respect to the dynamical scale $\Lambda$ 
and define the $k$-instanton part of the partition function $Z_k$:
\begin{align}
Z^{\textrm{inst}}(\vec{a},\Lambda,\epsilon_1,\epsilon_2)
=\sum_{k}{\Lambda^{2N_c k}}\,Z_{\,k}(\vec{a},\epsilon_1,\epsilon_2).
\end{align}
Let us $1$ and $2$-instanton partition functions for example.
\subsection*{1-instanton}
Terms with $|\vec {Y}|=1$ contribute to $1$-instanton part of the Nekrasov partition function (\ref{Nekrasov}).
Such Young diagrams take the form of $\vec{Y}=(\fundl\,,\phi,\phi,\cdots),\, (\phi, \fundl\,,\phi,\cdots),\, \cdots$.
Let us compute the factor (\ref{denominator}) for $\vec{Y}=(\fundl\,,\phi,\phi,\cdots)$:
\begin{align}
\prod_{\alpha,\beta=1}^{N_c}n_{\alpha,\beta}^{ (\sfund\,,\phi,\phi,\cdots)    }(\vec{a},\epsilon_1,\epsilon_2) 
&=n_{1,1}^{\vec{Y}} \cdot n_{1,2}^{\vec{Y}} \cdot n_{2,1}^{\vec{Y}} \cdot n _{1,3}^{\vec{Y}} \cdot n_{3,1}^{\vec{Y}}\cdots\nonumber\\
&=\epsilon_1\epsilon_2   \prod_{\alpha=2}^{N_c}  a_{\alpha ,1} ( a_{1,\alpha}+\epsilon).
\end{align}
Thus we get a $1$-instanton Nekrasov partition function as follows:
\begin{align}
\label{Nekrasov1}
Z_{\,k=1}(\vec{a},\epsilon_1,\epsilon_2)
=\sum_{\beta} \frac{1}{\epsilon_1\epsilon_2   \prod_{\alpha \neq \beta}^{N_c}  a_{\alpha ,\beta} ( a_{\beta,\alpha}+\epsilon)}
\end{align}
For $N_c=3$, the 1-instanton partition function (\ref{Nekrasov1}) becomes
\begin{align}
\label{SU3-1inst}
&Z_{\,k=1}({a_1},a_2,\epsilon_1,\epsilon_2)\nonumber\\
&\qquad=\frac{6({a_1}^2+{a_2}^2+a_1a_2-\epsilon^2)}
{\epsilon_1\epsilon_2(2a_1+a_2-\epsilon)(2a_1+a_2+\epsilon)(a_1+2a_2-\epsilon)(a_1+2a_2+\epsilon)(a_1-a_2-\epsilon)(a_1-a_2+\epsilon)},
\end{align}
where we use $a_1+a_2+a_3=0$.

\subsection*{ 2-instanton}
The Young diagrams which satisfy $|\vec{Y}|=2$  contribute to the 2-instanton partiton function.
There are three types of such Young diagrams:
$\vec{Y}=(\fundl\,,\fundl\,,\phi,\cdots)$, $(\anti\,,\phi,\phi,\cdots)$, $(\symm,\phi,\phi,\cdots) \cdots$.
Let us compute the contribution from $\vec{Y}=(\fundl\,,\fundl\,,\phi,\cdots)$.
\begin{align}
\prod_{\alpha,\beta=1}^{N_c}n_{\alpha,\beta}^{ (\sfund\,,\sfund\,,\phi,\cdots)    }(\vec{a},\epsilon_1,\epsilon_2) 
&=n_{1,1}^{\vec{Y}} \cdot n_{1,2}^{\vec{Y}} \cdot n_{2,1}^{\vec{Y}} \cdot n _{2,2}^{\vec{Y}}\nonumber\\
&\qquad\qquad \times n _{1,3}^{\vec{Y}} \cdot n_{3,1}^{\vec{Y}}\cdot  n _{1,4}^{\vec{Y}} \cdot n_{4,1}^{\vec{Y}}\cdots
n _{2,3}^{\vec{Y}} \cdot n_{3,2}^{\vec{Y}}\cdot  n _{2,4}^{\vec{Y}} \cdot n_{4,2}^{\vec{Y}}\cdots\nonumber\\
&=(\epsilon_1\epsilon_2 )^2   ( a_{1,2}+\epsilon_1) ( a_{1,2}-\epsilon_1) ( a_{1,2}+\epsilon_2)( a_{1,2}-\epsilon_2)\nonumber\\
&\qquad\qquad
\times \prod_{\beta=1,2}\prod_{\alpha \neq 1,2}^{N_c}  a_{\alpha ,\beta} ( a_{\beta,\alpha}+\epsilon).
\end{align}
The contribution from $\vec{Y}=(\anti\,,\phi,\phi,\cdots)$ is given by
\begin{align}
\prod_{\alpha,\beta=1}^{N_c}n_{\alpha,\beta}^{ (\santi\,,\phi,\phi,\cdots)    }(\vec{a},\epsilon_1,\epsilon_2) 
&=n_{1,1}^{\vec{Y}} \cdot n_{1,2}^{\vec{Y}} \cdot n_{2,1}^{\vec{Y}} \cdot n _{1,3}^{\vec{Y}} \cdot n_{3,1}^{\vec{Y}}\cdots\nonumber\\
&=(2\epsilon_1{\epsilon_2}^2(\epsilon_1-\epsilon_2 )) 
 \prod_{\alpha \neq 1}^{N_c}  a_{1,\alpha } ( a_{1,\alpha}+\epsilon)(a_{1,\alpha } +\epsilon_2)( a_{1,\alpha}+\epsilon+\epsilon_2).
\end{align}
Finally, the Young diagram $\vec{Y}=(\symm,\phi,\phi,\cdots) \cdots$ gives
\begin{align}
\prod_{\alpha,\beta=1}^{N_c}n_{\alpha,\beta}^{ (\ssymm,\phi,\phi,\cdots)   }(\vec{a},\epsilon_1,\epsilon_2) 
&=n_{1,1}^{\vec{Y}} \cdot n_{1,2}^{\vec{Y}} \cdot n_{2,1}^{\vec{Y}} \cdot n _{1,3}^{\vec{Y}} \cdot n_{3,1}^{\vec{Y}}\cdots\nonumber\\
&=(2{\epsilon_1}^2{\epsilon_2}(\epsilon_2-\epsilon_1 ))   
 \prod_{\alpha \neq 1}^{N_c}  a_{1,\alpha } ( a_{1,\alpha}+\epsilon)(a_{1,\alpha } +\epsilon_1)( a_{1,\alpha}+\epsilon+\epsilon_1).
\end{align}
Thus we get the $2$-instanton partition function:
\begin{align}
\label{Nekrasov2}
Z_{\,k=2}(\vec{a},\epsilon_1,\epsilon_2)
&=\sum_{\alpha<\beta}
\frac{1}{(\epsilon_1\epsilon_2 )^2   ( a_{\alpha,\beta}+\epsilon_1) ( a_{\alpha,\beta}-\epsilon_1) 
( a_{\alpha,\beta}+\epsilon_2)( a_{\alpha,\beta}-\epsilon_2)
 \prod_{i=\alpha,\beta}\prod_{\gamma \neq \alpha,\beta}^{N_c}  a_{\gamma ,i} ( a_{i,\gamma}+\epsilon)}\nonumber\\
 &+\sum_{\alpha}
\frac{1}{
(2\epsilon_1{\epsilon_2}^2(\epsilon_1-\epsilon_2 )) 
 \prod_{\beta \neq \alpha}^{N_c}  a_{\alpha,\beta } 
 ( a_{\alpha,\beta}+\epsilon)(a_{\alpha,\beta } +\epsilon_2)( a_{\alpha,\beta}+\epsilon+\epsilon_2)}\nonumber\\
 &+\sum_{\alpha}
\frac{1}{(2{\epsilon_1}^2{\epsilon_2}(\epsilon_2-\epsilon_1 ))   
 \prod_{\beta \neq \alpha}^{N_c}  a_{\alpha,\alpha } ( a_{\alpha,\alpha}+\epsilon)
 (a_{\alpha,\alpha } +\epsilon_1)( a_{\alpha,\alpha}+\epsilon+\epsilon_1)}.
\end{align}
In Appendix.\ref{sec:check}, we will use it in order to check our proposal.

Here the Coulomb moduli $a_{\alpha}$ have the mass dimension one.
In the latter part of the articles, we scale these parameters, 
and then $a_{\alpha}$ denotes the dimensionless Coulomb moduli parameter:
\begin{align}
a_{\alpha} \to \hbar \, a_{\alpha}.
\end{align}

\subsection{$\mathcal{W}_3$-algebra and $SU(3)$ Nekrasov formula}
The basic symmetry of conformal field theories is the Virasoro algebra \cite{Belavin:1984vu}.
In this section we study the $\mathcal{W}_3$-algebra \cite{Zamolodchikov:1985wn},
which is an enlarged conformal algebra.
This algebra is composed of the spin-$2$  energy-momentum tensor $T(z)$ and spin-$3$ current $W(z)$.
Their Laurent coefficients $L_n$ and $W_n$ satisfy the following commutation relations:
\begin{align}
&[L_n, L_m]=(n-m)L_{n+m}+\frac{c}{12}(n^3-n)\delta_{n,-m},\\
&[L_n, W_m]=(2n-m)W_{n+m},\\
&[W_n, W_m]=\frac{9}{2}\Big[ \frac{c}{3\cdot 5!}(n^2-1)(n^2-4)\delta_{n,-m}
+\frac{16}{22+5c}(n-m)\Lambda_{n+m}\nonumber\\
&\qquad\qquad\qquad +(n-m) \left( \frac{(n+m+2)(n+m+3)}{15}-\frac{(n+2)(m+2)}{6} \right)L_{n+m}\Big].
\end{align}
$\Lambda_n$ is a composite operator
\begin{align}
\Lambda_n=\sum_{m\in\mathbb{Z}}:L_m L_{n-m}:+\frac{x_n}{5}L_n,
\end{align}
where the constants $x$ are defined by
\begin{align}
&x_{2l}=(1-l)(1+l),\\
&x_{2l+1}=(1-l)(12+l).
\end{align}
We parametrize the central charge as $c=2(1-12Q^2)$.

The Hilbert space is spanned by the following basis of the descendants of the primary operator $V_{\vec{\alpha}}$:
\begin{align}
 L_{-Y_L}  W_{-Y_W}| \Delta_{\vec{\alpha}}\rangle\sim L_{-Y_L}  W_{-Y_W}V_{\vec{\alpha}}(z).
\end{align}
In this way the descendants are labelled by the pair of the Young diagrams $\mathcal{Y}=\{Y_L, Y_W\}$.
The conformal dimension of the operator is given by
\begin{align}
 \Delta_{(\vec{\alpha},\mathcal{Y})}= \Delta_{\vec{\alpha}} +|\mathcal{Y}|
 = \Delta_{\vec{\alpha}} +|Y_L|+|Y_W| .
\end{align}
The conformal dimensions of the primary with $\vec{\alpha}=(\alpha,\beta)$ are
\begin{align}
&\Delta_{\vec{\alpha}}=\alpha^2+\beta^2-Q^2,\nonumber\\
\label{confdim}
&w_{\vec{\alpha}}=\sqrt{\frac{4}{4-15Q^2}}\alpha(\alpha^2-3\beta^2),\\
&D(\Delta)=\frac{4\Delta}{4-15Q^2}+\frac{3Q^2}{4-15Q^2},\nonumber
\end{align}
where
\begin{align}
L_{0}| \Delta_{\vec{\alpha}}\rangle=\Delta_{\vec{\alpha}}| \Delta_{\vec{\alpha}}\rangle,\quad
W_{0}| \Delta_{\vec{\alpha}}\rangle=w_{\vec{\alpha}}| \Delta_{\vec{\alpha}}\rangle.
\end{align}

Let us consider the Shapovalov matrix of the $\mathcal{W}_3$-algebra \cite{Mironov:2009dr}\cite{Mironov:2009by}.
It is the Gram matrix of the following type:
\begin{align}
Q_{\Delta}(\mathcal{Y};\mathcal{Y}^{\prime})=\langle \Delta |  W_{Y_W}L_{Y_L} \cdot L_{-Y_L}  W_{-Y_W}| \Delta\rangle.
\end{align}
We propose that the Nekrasov partition function for $SU(3)$ pure super Yang-Mills theory coincides with the following element
of the Shapovalov matrix of $W_3$-algebra:
\subsection*{Proposal 3.1}
\begin{align}
Z_{SU(3),\, k}(\vec{a},\epsilon_1,\epsilon_2)
=\left( \frac{27}{4\epsilon_1\epsilon_2+15\epsilon^2}\right)^k
Q_{\Delta}^{-1}(\phi, [1^k] ; \,\phi, [1^k] ).
\end{align}
In other words, our proposal becomes
\begin{align}
Z_{SU(3),\, k}(\vec{a},e_1,e_2)
=\left( \frac{-27}{4-15e^2}\right)^k
Q_{\Delta}^{-1}(\phi, [1^k] ; \,\phi, [1^k] ).
\end{align}
after the scaling of the Coulomb moduli $\vec{a}\to \hbar\, \vec{a}$.
Here $\Delta=\Delta_{\vec{\alpha}}$ and the identification of the parameters is \cite{Mironov:2009by}
\begin{align}
&\alpha=\frac{\sqrt{3}}{2}(a_1+a_2),\nonumber\\
\label{su3param}
&\beta=\frac{1}{2}(-a_1+a_2),\\
&Q=e.\nonumber
\end{align}

\subsection*{Check}
Let us check our conjecture for $1$-instanton.
The level-$1$ Shapovalov matrix is \cite{Mironov:2009dr}\cite{Mironov:2009by}
\begin{equation}
Q_{\Delta}(\mathcal{Y} ; \,\mathcal{Y}^{\prime})|_{|\mathcal{Y}|,\,|\mathcal{Y}^{\prime}|=1}
=
\begin{array}{c|cc}
& (\,\fundl\,,\phi )& (\phi,\fundl\,)
\\ \hline (\,\fundl\,,\phi ) & 2\Delta& 3w
\\ (\phi,\fundl\,)               &3w       & \frac{9D\Delta}{2} \end{array},
\end{equation}
where the label is $\mathcal{Y}=(Y_L,Y_W)$.
The inverse of the matrix is given by
\begin{align}
Q_{\Delta}^{-1}(\mathcal{Y} ; \,\mathcal{Y}^{\prime})|_{|\mathcal{Y}|,\,|\mathcal{Y}^{\prime}|=1}
=
\frac{1}{9(D\Delta^2-w^2)}
\left(\begin{array}{cc}\frac{9D\Delta}{2}  & -3w \\-3w & 2\Delta\end{array}\right),
\end{align}
and the component of our interest is
\begin{align}
\label{Qphi1}
Q_{\Delta}^{-1}(\phi,\,\fundl\, ; \,\phi,\,\fundl )
=
\frac{2\Delta}{9(D\Delta^2-w^2)}.
\end{align}
Let us evaluate the element (\ref{Qphi1}) under (\ref{su3param}).
By using (\ref{confdim}), we can factorize the determinant of the Shapovalov matrix \cite{Mironov:2009by}:
\begin{align}
D\Delta^2-w^2=
\frac{4}{4-15\,Q^2}\left( \beta^2-\frac{Q^2}{4}  \right)\left( (\beta-Q)^2-3\alpha^2\right)\left(  (\beta+Q)^2-3\alpha^2\right).
\end{align}
With (\ref{su3param}), we can rewrite it as a factor appearing in the Nekrasov partition function:
\begin{align}
D\Delta^2-w^2=
\frac{1}{4-15\,e^2}
(a_{12}-e)(a_{12}+e)(a_{23}-e)(a_{23}+e)(a_{31}-e)(a_{31}+e).
\end{align}
Similarly, the conformal dimension becomes
\begin{align}
\Delta=a_1^2+a_2^2+a_1a_2-e^2.
\end{align}
Thus we can rewrite (\ref{Qphi1}) as the $1$-instanton partition function (\ref{SU3-1inst}):
\begin{align}
Z_{SU(3),\, k=1}(\vec{a},e_1,e_2)
&=\frac{-6}{4-15\,e^2}\frac{\Delta}{D\Delta^2-w^2}\nonumber\\
&=\frac{-27}{4-15\,e^2}\,Q_{\Delta}^{-1}(\phi, \fundl \,; \phi, \fundl \,).
\end{align}
In this way, we can verify the conjecture explicitly for $1$-instanton.
We verify it for $2$-instanton in Appendix.\ref{sec:check}.

\subsection{Derivation from AGT conjecture}
\subsubsection{AGT-W relation for $N_c=3$, $N_f=6$ gauge theory}
AGT-W conjecture \cite{Wyllard:2009hg} gives relation between 
$SU(N_c)$ $N_f=2N_c$ gauge theory and the conformal Toda theory with $\mathcal{W}_N$-symmetry.
The core of the claim for $N_c=3$, $N_f=6$ gauge theory is that the Nekrasov partition function is equal to
the 4-point spherical conformal block of the $\mathcal{W}_3$-algebra \cite{Mironov:2009by}:
\begin{align}
Z^{\textrm{inst}}_{\,SU(3)}(a_1,a_2,\vec{\mu}, x,e_1,e_2)=
\mathcal{B}_{\Delta_{\vec{\alpha}}}
\tiny{\left[ {\begin{array}{cc}\Delta_1 \hspace{-0.2cm}& \Delta_2 \\ \Delta_3 \hspace{-0.2cm}& \Delta_4 \end{array}}  \right]}
(x).
\end{align}
Here $\mathcal{B}$ is the 4-point conformal block of the $\mathcal{W}_3$-algebra
\begin{align}
\label{4pcb}
\mathcal{B}_{\Delta_{\vec{\alpha}}}
\tiny{\left[ {\begin{array}{cc}\Delta_1 \hspace{-0.2cm}& \Delta_2 \\ \Delta_3 \hspace{-0.2cm}& \Delta_4 \end{array}}  \right]}(x)=
\sum_{\mathcal{Y},\mathcal{Y}^{\prime} \hspace{0.1cm} |  \hspace{0.1cm}   |\mathcal{Y}|=|\mathcal{Y}^{\prime}|}
x^{|\mathcal{Y}|}\,
\bar{\rho}_{\alpha_1\alpha_2; \alpha     }(\mathcal{Y})\,
Q_{\Delta}^{-1}(\mathcal{Y};\mathcal{Y}^{\prime})\,
{\rho}_{\alpha\alpha_3 \alpha_4 }(\mathcal{Y}^{\prime}).
\end{align}
See Appendix.\ref{sec:cb} for the definition of it.
Recall that the theory is conformal and  the partition function does not depend on the overall scale $\hbar$:
\begin{align}
Z^{\textrm{inst}}_{\,SU(3)}(\vec{a}\hbar,\,\vec{\mu}\hbar, \,x,\,e_1\hbar,\,e_2\hbar)=Z^{\textrm{inst}}_{\,SU(3)}(\vec{a},\vec{\mu}, x,e_1,e_2).
\end{align}
Thus we have to identify these parameters of the gauge theory with the variables of the conformal block.
The identification between the internal momentum $\vec{\alpha}$ and the Coulomb moduli $\vec{a}$ is the same as the above.
The relation between the masses and the external momentum is \cite{Mironov:2009by}
\begin{align}
&\mu_1=\frac{2}{\sqrt{3}}\alpha_2+f(\alpha_1,\beta_1),  
\qquad\quad \mu_ 4=\frac{2}{\sqrt{3}}\alpha_4+f(\alpha_3,\beta_3)\nonumber\\
&\mu_2=\frac{-1}{\sqrt{3}}\alpha_2-\beta_2+f(\alpha_1,\beta_1),
 \quad \mu_ 5=\frac{-1}{\sqrt{3}}\alpha_4-\beta_4+f(\alpha_3,\beta_3)\\
&\mu_3=\frac{-1}{\sqrt{3}}\alpha_2+\beta_2+f(\alpha_1,\beta_1),  
\quad \mu_ 6=\frac{-1}{\sqrt{3}}\alpha_4+\beta_4+f(\alpha_3,\beta_3)\nonumber
\end{align}
There exist $2\times4=8$ external momentum in the $\mathcal{W}$-algebra side, on the other hand
the gauge theory has $6$ mass parameters.
This means $2$ parameters of the $\mathcal{W}$-algebra are redundant for AGT correspondence.
In \cite{Wyllard:2009hg} Wyllard therefore proposed that we should make two external states $\mathcal{W}$-null \cite{Kanno:2009ga}.
Then we have $3\times 3=9$ choices of such a state.
Now we choose a $\mathcal{W}$-null vector $\vec{\alpha}=(\alpha,-Q/2)$ for simplicity.
Then the factor $f$ is given by \cite{Mironov:2009by}
\begin{align}
f(\alpha_1,-\epsilon/2)=\frac{-1}{\sqrt{3}}\alpha_1+\frac{Q}{2},\quad
f(\alpha_3,-\epsilon/2)=\frac{1}{\sqrt{3}}\alpha_3+\frac{Q}{2}.
\end{align}
Then we finfd
\begin{align}
\label{massprod1}
&\mu_1\mu_2\mu_3=\frac{2}{3\sqrt{3}}\alpha_2({\alpha_2}^2-3{\beta_2}^2)-f(\vec{\alpha_1})({\alpha_2}^2+{\beta_2}^2)+{f(\vec{\alpha_1})}^3\\
\label{massprod2}
&\mu_4\mu_5\mu_6=\frac{2}{3\sqrt{3}}\alpha_4({\alpha_4}^2-3{\beta_4}^2)-f(\vec{\alpha_3})({\alpha_4}^2+{\beta_4}^2)+{f(\vec{\alpha_3})}^3.
\end{align}
We will use the relations (\ref{massprod1}), (\ref{massprod2}) later.

\subsubsection{decoupling limit of massive hypermultiplets}
We consider the following decoupling limit of the massive hypermultiplets:
\begin{align}
&x\to 0,\nonumber\\
\label{scalinglim}
&\mu_i\to\infty,\\
&x\prod_{i=1}^{6}\mu_i=\Lambda.\nonumber
\end{align}
This limit makes the external momentum infinity:
\begin{align}
\alpha_1, \,\alpha_2,\,\beta_2,\, \alpha_3,\,\alpha_4, \,\beta_4 \to \infty.
\end{align}
The resulting theory is the $SU(3)$ pure super Yang-Mills theory,
and thereby we can "derive" our proposal from the AGT-W conjecture by using this scaling limit.
\subsubsection*{recursion relations and asymptotic behavior}
The 4-point conformal block consists of the 3-point spherical conformal blocks and the Shapovalov matrix (Appendix.\ref{sec:cb}).
Then we have to study the asymptotic behavior of the 3-point spherical conformal blocks in order to
evaluate the scaling limit of the 4-point conformal block.
For the purpose, we show the fact that among correlators $\langle V_{\alpha, \mathcal{Y}}| V_1(1)V_2(0)\rangle$ 
and $\langle V_{\alpha, \mathcal{Y}}(0) V_3(1)V_4(\infty)\rangle$
with fixed $|\mathcal{Y}|\equiv|Y_L|+|Y_W|=n$,
the 3-point blocks with $\mathcal{Y}=(\phi,[1^n])$ give the leading contribution in the limit:
\begin{align}
&\langle L_{-\phi}W_{[1^n]} V_{\alpha}| V_1(1)V_2(0)\rangle=
\left(2w_1-w_2-\frac{3w_1}{2\Delta_1}(\Delta_{1}+\Delta_{2})\right)^n
\langle V_{\alpha}| V_1(1)V_2(0)\rangle\nonumber\\
&\hspace{10cm}
+\textrm{ sub-leading terms},\\
&\langle (L_{-\phi}W_{[1^n]} V_{\alpha})(0) V_3(1)V_4(\infty)\rangle=
\left(w_3+w_4-\frac{3w_3}{2\Delta_3}(\Delta_{3}-\Delta_{4})\right)^n
\langle V_{\alpha}(0) V_1(1)V_2(\infty)\rangle\nonumber\\
&\hspace{10cm}+\textrm{ sub-leading terms}.
\end{align}
This means that in the limit (\ref{scalinglim}) only the following 3-point conformal blocks
show the dominant behavior and
can contribute to the 4-point conformal block  (\ref{4pcb}):
\begin{align}
\label{scalingbr}
&\bar{\rho}_{\alpha_1\alpha_2;\alpha}(\phi,[1^n])=
\left(2w_1-w_2-\frac{3w_1}{2\Delta_1}(\Delta_{1}+\Delta_{2})\right)^n
+\textrm{ sub-leading terms},\\
\label{scalingr}
&{\rho}_{\alpha\alpha_3\alpha_4}(\phi,[1^n])=
\left(w_3+w_4-\frac{3w_3}{2\Delta_3}(\Delta_{3}-\Delta_{4})\right)^n
+\textrm{ sub-leading terms}.
\end{align}

Let us prove the above statement.
The key is the Mironov-Mironov-Morozov-Morozov recursion relations \cite{Mironov:2009dr} for a descendant $ V_{\textbf{a}}$:
\begin{align}
\label{recursionL1}
&\langle L_{-n} V_{\textbf{a}}| V_1(1)V_2(0)\rangle
=(\Delta_{\textbf{a} } +n\Delta_{1}-\Delta_{2})
\langle V_{\textbf{a}}| V_1(1)V_2(0) \rangle,\\
\label{recursionL2}
&\langle (L_{-n} V_{\textbf{a}})(0) V_3(1)V_4(\infty)\rangle
=(\Delta_{\textbf{a} } +n\Delta_{3}-\Delta_{4})
\langle V_{\textbf{a}}(0) V_3(1)V_4(\infty) \rangle,\\
\label{recursionW1}
&\langle W_{-n} V_{\textbf{a}}| V_1(1)V_2(0)\rangle
=\langle W_0V_{\textbf{a}}| V_1(1)V_2(0) \rangle
+\left(\frac{n(n+3)w_{1}}{2}-w_{2}\right)
\langle V_{\textbf{a}}| V_1(1)V_2(0) \rangle\nonumber\\
&\hspace{10cm} +n\langle V_{\textbf{a}}| (W_{-1}V_1)(1)V_2(0) \rangle,\\
\label{recursionW2}
&\langle (W_{-n} V_{\textbf{a}})(0) V_3(1)V_4(\infty)\rangle
=\langle (W_0V_{\textbf{a}})(0) V_3(1)V_4(\infty) \rangle
+\left(\frac{n(3-n)w_{3}}{2}+w_{4}\right)
\langle V_{\textbf{a}}(0) V_3(1)V_4(\infty) \rangle\nonumber\\
&\hspace{10cm} +n\langle V_{\textbf{a}}(0) (W_{-1}V_3)(1)V_4(\infty) \rangle.
\end{align}

We now study a pair of partitions $\mathcal{Y}\equiv\{Y_L,Y_W\}$ with fixed number of boxes $n=|Y_L|+|Y_W|$.
Let us show the fact that the partition $\mathcal{Y}$ whose correlator $\langle V_{\alpha, \mathcal{Y}}| V_1(1)V_2(0)\rangle$  
dominates in the limit  is in the shape of $\mathcal{Y}\equiv\{\phi,Y_W\}$.
We can prove it by inductive argument.
First we compare
$\langle L_{-m} V_{\textbf{a}}| V_1(1)V_2(0)\rangle$ and $\langle W_{-m} V_{\textbf{a}}| V_1(1)V_2(0)\rangle$.
In the right hand sides of the recursion relations (\ref{recursionL1}) and (\ref{recursionW1}),
$L_{-m}$ creates a factor $\Delta_{i}\sim{\alpha_{i}}^2$
and $W_{-m}$ gives $w_{i}\sim{\alpha_{i}}^3$ in  the limit.
Moreover the first term of (\ref{recursionW1})
\begin{align}
\langle W_0V_{\textbf{a}}| V_1(1)V_2(0) \rangle
\end{align}
can not contribute to the dominant behavior, since the action of $W_0$ gives
\begin{align}
| W_0V_{\alpha,\mathcal{Y}}\rangle=\sum_{|\mathcal{Y}^{\prime}|=|\mathcal{Y}|}c(\alpha)\,| V_{\alpha,\mathcal{Y}^{\prime}}\rangle.
\end{align}
As we will show below, the third term of the right hand side of the recursion relation (\ref{recursionW1}) is also negligible
in the limit.
The correlator $\langle W_{-m} V_{\textbf{a}}| V_1(1)V_2(0)\rangle$ is hence the dominant one.
By using this result inductively, the correlator $\langle L_{-Y_L}W_{-Y_W} V_{\alpha}| V_1(1)V_2(0)\rangle$
become dominant for $Y_L=\phi$ and $Y_W=\left[1^n\right]$.

Next, let us evaluate the leading correlator $\langle W_{-1}^n V_{\alpha}| V_1(1)V_2(0)\rangle$.
Since we choose the external state $V_1$ as the $\mathcal{W}$-null \cite{Wyllard:2009hg}\cite{Mironov:2009dr}, 
the action of the $W_{-1}$ operator on the primary is
$W_{-1}V_1=\frac{3w_1}{2\Delta_1}L_{-1}V_1$.
Then we obtain
\begin{align}
\label{3rdterm}
\langle V_{\textbf{a}}| (W_{-1}V_1)(1)V_2(0) \rangle
&=\frac{3w_1}{2\Delta_1}
\langle V_{\textbf{a}}| (L_{-1}V_1)(1)V_2(0) \rangle\nonumber\\
&=\frac{3w_1}{2\Delta}
(\Delta_{\textbf{a}}-\Delta_1-\Delta_2)
\langle V_{\textbf{a}}| V_1(1)V_2(0) \rangle
.
\end{align}
Here we have used the relation
\begin{align}
\langle V_{\textbf{a}}| (L_{-1}V_1)(1)V_2(0) \rangle\
=(\Delta_{\textbf{a}}-\Delta_1-\Delta_2)
\langle V_{\textbf{a}}| V_1(1)V_2(0) \rangle.
\end{align}
The conformal dimension is 
\begin{align}
\Delta_{\textbf{a}}=
|Y_L|+|Y_W|+\Delta_{\alpha}
\end{align}
for $V_{\textbf{a}}=L_{-Y_L}W_{-Y_W}V_{\alpha}$, since 
$
L_0L_{-Y_L}W_{-Y_W}V_{\alpha}
=(|Y_L|+|Y_W|+\Delta_{\alpha})
V_{\alpha}
$ holds.
This factor is negligible in the limit $\alpha_i,\,\beta_i\to\infty$.
Therefore the term (\ref{3rdterm}) behaves as $-\Delta_1-\Delta_2$ in the decoupling limit,
and the second term of the right hand side of the recursion relation (\ref{recursionW1}) is dominant in the limit.
Thus we obtain the leading term of the 3-point function:
\begin{align}
\langle W_{[1^n]} V_{\alpha}| V_1(1)V_2(0)\rangle
=\left(2w_{1}-w_{2} -\frac{3w_1}{2\Delta_1}(\Delta_1+\Delta_2)\right)^n
\langle V_{\alpha}| V_1(1)V_2(0) \rangle+\cdots.
\end{align}
Similar argument shows that the following correlator is also leading:
\begin{align}
\langle (W_{[1^n]} V_{\alpha})(0) V_3(1)V_4(\infty)\rangle=
\left(w_3+w_4-\frac{3w_3}{2\Delta_3}(\Delta_{3}-\Delta_{4})\right)^n
\langle V_{\alpha}(0) V_1(1)V_2(\infty)\rangle+\cdots.
\end{align}

\subsubsection*{scaling limit of the conformal block}
In this paper we choose a specific W-null state $\vec{\alpha}_1=(\alpha_1, -\frac{\epsilon}{2})$ for simplicity.
The argument does not change when we choose any other null state.
The dimensions of the state with $\vec{\alpha}_1$ are
\begin{align}
&\Delta_1={\alpha_1}^2-\frac{3}{4}Q^2,\nonumber\\
&{w_1}=\frac{2{\alpha_1} {\Delta_1}}{\sqrt{4-15Q^2}},\\
&D_1=\frac{4{\alpha_1}^2}{4-15Q^2}.\nonumber
\end{align}
Then the factor appearing in (\ref{scalingbr}) becomes
\begin{align}
2w_1-w_2-\frac{3w_1}{2\Delta_1}(\Delta_1+\Delta_2)
&\nonumber\\
&\hspace{-3cm}=\frac{-3\sqrt{3}}{\sqrt{4-15Q^2}}\left( 
\frac{\sqrt{4-15Q^2}}{3\sqrt{3}}w_2-
\left( -\frac{\alpha_1}{\sqrt{3}}\right)({\alpha_2}^2+{\beta_2}^2)
+\left( -\frac{\alpha_1}{\sqrt{3}}\right)^3
  \right)
  -\frac{9\alpha_1Q^2}{4\sqrt{4-15Q^2}}\nonumber\\
  &\hspace{-3cm}=\frac{-3\sqrt{3}}{\sqrt{4-15Q^2}}\mu_1\mu_2\mu_3- \frac{9\alpha_1Q^2}{4\sqrt{4-15Q^2}}.
\end{align}
Here we use (\ref{massprod1}).
We can also find the similar relation for (\ref{scalingr}):
\begin{align}
w_3+w_4+\frac{3w_3}{2\Delta_3}(\Delta_3-\Delta_4)
=\frac{3\sqrt{3}}{\sqrt{4-15Q^2}}\mu_4\mu_5\mu_6+ \mathcal{O}(\alpha_3).
\end{align}
Thus we get the following relation in the limit (\ref{scalinglim}):
\begin{align}
\lim \left( 2w_1-w_2-\frac{3w_1}{2\Delta_1}(\Delta_1+\Delta_2)\right)
\left( w_3+w_4+\frac{3w_3}{2\Delta_3}(\Delta_3-\Delta_4) \right)x=-\frac{27}{4-15Q^2}\Lambda^6.
\end{align}
Recall that for fixed $|\mathcal{Y}|=n$ the dominant contribution of the sum (\ref{4pcb}) 
comes from (\ref{scalingbr}) and (\ref{scalingr}).
Therefore we get the following result:
\begin{align}
\lim \left[x^{|\mathcal{Y}|}\, \bar{\rho}_{\alpha_1 \alpha_2 ; \alpha} ({\mathcal{Y}})\,
{\rho}_{\alpha \alpha_3  \alpha_4} ( \mathcal{Y}^{\prime} )\right]_{|\mathcal{Y}|=|\mathcal{Y}^{\prime}|=n}
=\delta_{\mathcal{Y},\,(\phi, [1^n])}\delta_{\mathcal{Y}^{\prime},\,(\phi, [1^n])}
\left(\frac{-27}{4-15Q^2}\right)^n\Lambda^{6n}
\end{align}
Therefore only the terms with $\mathcal{Y}=(\phi, [1^n])$ survive in the limit.
This means that the scaling limit of the conformal block is given by the Shapovalov matrix:
\begin{align}
\lim \mathcal{B} = \sum_n \left( \frac{-27}{4-15Q^2}\Lambda^6 \right)^n Q_{\Delta}^{-1}(\phi, [1^n]; \phi, [1^n]).
\end{align}
The gauge theory with $6$-flavors becomes $SU(3)$ pure Yang-Mills theory in the decoupling limit.
By assuming the AGT-W conjecture \cite{Wyllard:2009hg}\cite{Mironov:2009by},
 we can thus prove our proposal for $SU(3)$ pure Yang-Mills theory
\begin{align}
Z^{\textrm{inst}}_{\,SU(3)}
= \sum_n \left( \frac{-27}{4-15Q^2}\Lambda^6 \right)^n Q_{\Delta}^{-1}(\phi, [1^n]; \phi, [1^n]).
\end{align}

\section{Conclusion}
In this paper we have proposed a relation between representation theory of the $\mathcal{W}_3$-symmetry
and the instanton counting of $SU(3)$ pure super Yang-Mills theory.
We found that the Nekrasov partition function for the Yang-Mills theory is equal to
the following elements of the inverse Shapovalov matrix:
\begin{align}
\label{result}
Z^{\textrm{inst}}_{\,SU(3)}(a,b,\Lambda,e_1,e_2)
= 1+\sum_{n=1}^{\infty} \left( \frac{-27}{4-15e^2}\Lambda^6 \right)^n Q_{\Delta}^{-1}(\phi, [1^n]; \phi, [1^n]).
\end{align}
We also proved our proposal by assuming the AGT-W conjecture and taking the
decoupling limit of massive hypermultiplets.
Then the asymptotic behavior of the 3-point spherical conformal blocks (\ref{scalingbr})(\ref{scalingr})
played a key role.
The proposal (\ref{result}) is a skeleton of the AGT-W conjecture.
Thus the study of our proposal will give an efficient check of the original AGT-W conjecture.
In this paper we verified our proposal explicitly up to $2$-instanton.

Our proposal is a simple and nontrivial extension of the Gaiotto-Marshakov-Mironov proposal 
\cite{Gaiotto:2009ma}\cite{Marshakov:2009gn},
and it suggests that there exists a direct connection between instanton counting
and conformal symmetry.
Then it is very natural to expect that the Nekrasov partition function for $SU(N)$ pure Yang-Mills theory
to be related to the Shapovalov matrix of the $\mathcal{W}_N$-symmetry.
It would be very nice to find the explicit relation between them.
We also expect that we can recast the theory of instanton
 in the language of the representation theory of conformal symmetry and vice versa.
It would deepen our understanding of nonperturbative dynamics of $\mathcal{N}=2$ theories and
2-dimensional CFT's.

In this paper we focused on the pure Yang-Mills theory.
It is possible to extend our proposal for $SU(3)$ gauge theories with $N_f=1,2,\cdots,5$ flavors
by studying an appropriate decoupling limit of hypermultiplets.
Moreover multi-point conformal blocks for the Virasoro algebra are related to the Nekrasov
partition functions of $SU(2)$ quiver gauge theories \cite{Alba:2009ya}.
The $SU(3)$ quiver theories would therefore be related to $\mathcal{W}_3$-algebra.
Matrix models \cite{Dijkgraaf:2009pc}\cite{Itoyama:2009sc}\cite{Eguchi:2009gf}\cite{Schiappa:2009cc}\cite{Fujita:2009gf}
 may give an effective tool to study these extended relations.

In \cite{Awata:2009ur}, an analog of the AGT conjecture in the 5-dimensional $SU(2)$ gauge theory was found.
The $q$-deformed Virasoro algebra is the conformal symmetry which plays a role in the extended  AGT correspondence.
We would be able to connect 
the representation theory of the deformed $\mathcal{W}$-algebra and
the 5-dimensional Nekrasov partition functions.

\section{Acknowledgement}
The auther is supported by JSPS Grant-in-Aid for Creative Scientific Research, No.19GS0219.

\appendix

\section*{Appendix}
\section{Conformal Symmetry and Conformal Blocks}
\label{sec:cb}
\subsection{structure of correlators}
In this appendix we summarize the definition of the 4-point conformal block.
See \cite{Marshakov:2009gs}\cite{Mironov:2009dr}\cite{Mironov:2009by}  for details.

The basic objects in conformal field theory are the states (vertex operators $V_{\textbf{a}}$), the norms (2-point functions) and
the structure constants (OPE).

Let us consider the contravariant  form (the Shapovalov form) on the Verma module 
for the representation of a certain conformal symmetry.
The 2-point functions are characterized by the following type of Gram matrix  
\begin{align}
K_{ \textbf{a} \textbf{b}}
=\langle  V_{\textbf{a}} | V_{\textbf{b}}  \rangle.
\end{align}
Its matrix element is nonzero if and only if the conformal dimensions are equal $\Delta_{\textbf{a}}=\Delta_{\textbf{b}}$.
As we will see, the so-called Shapovalov matrix is the model independent part of the matrix which 
depends only on the representation theory of conformal symmetry.
The structure constants of the operator algebra are encoded in the OPE of the vertex operators:
\begin{align}
V_{\textbf{a}}(z) V_{ \textbf{b}}(z^{\prime})=
\sum_{\textbf{c}}
\frac{C^{\textbf{c}}_{\textbf{a}\textbf{b}} V_{ \textbf{c}}(z^{\prime})}
{{(z-z^{\prime})}^{   \Delta_{\textbf{a}}+\Delta_{\textbf{b}}-\Delta_{\textbf{c}}   }}.
\end{align}

Let us consider the following 3 point functions:
\begin{align}
\label{gammabar}
\bar{\Gamma}_{\textbf{a}_1 \textbf{a}_2 ; \textbf{a}}
&\equiv
\langle V_{\textbf{a}} | V_{\textbf{a}_1}(1) V_{ \textbf{a}_2}(0)  \rangle,\nonumber\\
&=\sum_{\textbf{b}} C^{\textbf{b}}_{\textbf{a}_1\textbf{a}_2} K_{ \textbf{a} \textbf{b}}\\
{\Gamma}_{\textbf{b} \textbf{a}_3 \textbf{a}_4 }
&\equiv
\langle V_{\textbf{b}}(0)  V_{\textbf{a}_3}(1) V_{ \textbf{a}_4}(\infty)  \rangle.
\end{align}
As we will see, they are basic building blocks of the correlators.

In this articles, we are interested in the 4-point spherical correlation functions. 
\begin{align}
G^{(4)}
\tiny{\left[ {\begin{array}{cc}\Delta_1 \hspace{-0.2cm}& \Delta_2 \\ \Delta_3 \hspace{-0.2cm}& \Delta_4 \end{array}}  \right]}
(x)
&\equiv
\langle   V_{\textbf{a}_1}(x) V_{ \textbf{a}_2}(0) V_{\textbf{a}_3}(1) V_{ \textbf{a}_4}(\infty)  \rangle\nonumber\\
&=\sum_{\textbf{b}}
x^{ -\Delta_1 - \Delta_2 }\,x^{ \Delta_{\textbf{b} }}\,
 C^{\textbf{b}}_{\textbf{a}_1\textbf{a}_2}\, {\Gamma}_{\textbf{b} \textbf{a}_3 \textbf{a}_4 }.
\end{align}
By using  (\ref{gammabar}), we find
\begin{align}
G^{(4)}
\tiny{\left[ {\begin{array}{cc}\Delta_1 \hspace{-0.2cm}& \Delta_2 \\ \Delta_3 \hspace{-0.2cm}& \Delta_4 \end{array}}  \right]}
(x)
=\sum_{\textbf{b}}
x^{ -\Delta_1 - \Delta_2 }\,x^{ \Delta_{\textbf{b} }}\,
\bar{\Gamma}_{\textbf{a}_1 \textbf{a}_2 ; \textbf{a}}\,
{(K^{-1})}^{\textbf{a} \textbf{b}}\,
 {\Gamma}_{\textbf{b} \textbf{a}_3 \textbf{a}_4 }.
\end{align}

Since we consider the correlation functions of the primaries, let us assume $V_{\textbf{a}_i}$ are primary $V_{\alpha_i}$.
We introduce the label of the descendants $ \mathcal{Y}$ .
Then $\Gamma$ and $\bar{\Gamma}$ are 
\begin{align}
& \bar{\Gamma}_{\alpha_1 \alpha_2 ; (\alpha, \mathcal{Y})}=
 \bar{\rho}_{\alpha_1 \alpha_2 ; \alpha} (\mathcal{Y})\, C_{\alpha_1 \alpha_2 ; \alpha},\\
&{\Gamma}_{(\beta, \mathcal{Y})\alpha_3 \alpha_4 }=
{\rho}_{\beta \alpha_3  \alpha_4} (\mathcal{Y})\, C_{\beta \alpha_3  \alpha_4}.
\end{align}
Here $\bar{\rho}$ and $\rho$ are 3-point shperical conformal blocks, 
which are objects of the representation theory of conformal algebra.
Meanwhile, the 3-point functions for primaries
\begin{align}
&C_{\alpha_1 \alpha_2 ; \alpha}\equiv\bar{\Gamma}_{\alpha_1 \alpha_2 ; (\alpha,\phi)},\\
&C_{\beta \alpha_3  \alpha_4}\equiv{\Gamma}_{(\beta, \phi)\alpha_3 \alpha_4 },
\end{align}
depend on the choice of a CFT model.

\subsection{conformal block of Virasoro algebra}
We take the Virasoro symmetry for example.
The Virasoro algebra $\mathcal{W}_2$ is an infinite dimensional symmetry of
2-dimensional CFT models.
It is generated by the energy-momentum tensor $T(z)=\sum z^{-n-2}L_n$.
The symmetry take the following form:
\begin{align}
[L_n,L_m]=(n-m)L_{n+m}+\frac{c}{12}n(n^2-1)\delta_{n,-m}.
\end{align}
The action of these generators on a primary operator $V_{\alpha}(z)$ gives descendants, 
which  are labeled by the Young diagrams $Y=\{ Y_1,\, Y_2,\, \cdots \}$:
\begin{align}
V_{\alpha,\, Y}&=L_{-Y}V_{\alpha}\nonumber\\
&=L_{-L_l}\cdots L_{-L_2}L_{-Y_1}V_{\alpha}.
\end{align}
The conformal dimension of the descendant is $\Delta_{\alpha,\, Y}=\Delta_{\alpha}+|Y|$.
The Gram matrix for the Verma module of the Virasoro symmetry is
\begin{align}
K_{ (\alpha, Y),\, (\beta,Y^{\prime})}
&=\delta_{|Y|,|Y^{\prime}|}K_{\alpha\beta}Q_{\alpha\beta}(Y,Y^{\prime})\nonumber\\
&=\delta_{|Y|,|Y^{\prime}|}\delta_{\alpha,\beta}K_{\alpha}Q_{\Delta_{\alpha}}(Y,Y^{\prime}).
\end{align}
Here the norms of primaries characterize the dependence of the norms on a CFT model.
The Shapovalov matrix $Q_{\Delta_{\alpha}}(Y,Y^{\prime})$ is independent of a choice of a model.

In the case of the Virasoro symmetry, the model-dependent part of the 3-point functions also possesses a peculiar feature. 
The Virasoro symmetry implies that the two conformal blocks are equal \cite{Mironov:2009dr}: 
\begin{align}
 \bar{\rho}_{\alpha_1 \alpha_2 ; \alpha} ({Y})
 ={\rho}_{\alpha \alpha_1  \alpha_2} ({Y}).
\end{align}
This property simplify computation of the Virasoro conformal blocks.

Then the spherical 4-point function becomes
\begin{align}
G^{(4)}
\tiny{\left[ {\begin{array}{cc}\Delta_1 \hspace{-0.2cm}& \Delta_2 \\ \Delta_3 \hspace{-0.2cm}& \Delta_4 \end{array}}  \right]}
(x)
=\sum_{\alpha}\,
x^{ \Delta_{\alpha}}\,
(C_{\alpha_1 \alpha_2 ; \alpha}\, (K^{-1})^{\alpha}\, C_{\alpha \alpha_3  \alpha_4})\,
\mathcal{F}_{\Delta_{\alpha}}
{\left[ {\begin{array}{cc}\Delta_1 \hspace{-0.2cm}& \Delta_2 \\ \Delta_3 \hspace{-0.2cm}& \Delta_4 \end{array}}  \right]}(x).
\end{align}
Here $\mathcal{F}$ is the model-independent part of the 4-point function.
It is given by the conformal block $\mathcal{B}$
\begin{align}
\mathcal{F}_{\Delta_{\alpha}}
\tiny{\left[ {\begin{array}{cc}\Delta_1 \hspace{-0.2cm}& \Delta_2 \\ \Delta_3 \hspace{-0.2cm}& \Delta_4 \end{array}}  \right]}(x)
=
x^{ -\Delta_1 - \Delta_2 }\,
\mathcal{B}_{\Delta_{\alpha}}
{\left[ {\begin{array}{cc}\Delta_1 \hspace{-0.2cm}& \Delta_2 \\ \Delta_3 \hspace{-0.2cm}& \Delta_4 \end{array}}  \right]}(x),
\end{align}
and the 4-point sherical  conformal block $\mathcal{B}$ is defined by
\begin{align}
\mathcal{B}_{\Delta_{\alpha}}
\tiny{\left[ {\begin{array}{cc}\Delta_1 \hspace{-0.2cm}& \Delta_2 \\ \Delta_3 \hspace{-0.2cm}& \Delta_4 \end{array}}  \right]}(x)
&=\sum_{|Y|=|Y^{\prime}|}\,
x^{|Y|}\, \bar{\rho}_{\alpha_1 \alpha_2 ; \alpha} ({Y})\,
Q_{\Delta_{\alpha}}^{-1}(Y,Y^{\prime})\,
{\rho}_{\alpha \alpha_3  \alpha_4} ( Y^{\prime} )\nonumber\\
&=1+\sum_{k=1}^{\infty}x^k\, 
\mathcal{F}_{\Delta_{\alpha}}^{(k)}
\tiny{\left[ {\begin{array}{cc}\Delta_1 \hspace{-0.2cm}& \Delta_2 \\ \Delta_3 \hspace{-0.2cm}& \Delta_4 \end{array}}  \right]}.
\end{align}
Notice that $ \bar{\rho}_{\alpha_1 \alpha_2 ; \alpha} ({Y})
 ={\rho}_{\alpha \alpha_1  \alpha_2} ({Y})$ holds for the Virasoro symmetry.
 The equivalence does not hold for the extended $\mathcal{W}$-symmetry. 
 
 Recall that the conformal block $\mathcal{B}$, which does not depend on a choice of a model, is equal to
 the Nekrasov partition function of $2N_c=N_f=4$ gauge theory in the AGT conjecture.
 This suggests that we can recast a Nekrasov partition function into an object of the representation theory of conformal symmetry.
 It is also conjectured that $C_{\alpha_1 \alpha_2 ; \alpha}\, (K^{-1})^{\alpha}\, C_{\alpha \alpha_3  \alpha_4}$ 
 gives the perturbative part of the Nekrasov partition function for a specific CFT model \cite{Alday:2009aq}.
 Thus the AGT conjecture implies that a full Nekrasov partition function relate to a physical correlator 
 of a certain CFT model.
 Notice that we does not study physical correlators but their holomorphic (chiral) part, 
 since we concentrate our attention on the instanton part of partition functions.

\section{Check of the Conjecture for Level-2}
\label{sec:check}
In this section, we verify our conjecture at 2-instanton level.
Let us recall the identification of parameters \cite{Mironov:2009by}:
\begin{align}
&c=2-24\, \left( e_{{1}}+e_{{2}} \right) ^{2},\\
&\Delta={a}^{2}+{b}^{2}+ab- \left( e_{{1}}+e_{{2}} \right) ^{2},\\
&w=\sqrt {\frac{27}{ 4-15 \left( e_{{1}}+e_{{2}} \right) ^{2
}}}
ab \left( a+b \right),\\
&D=4\,{\frac {{a}^{2}+{b}^{2}+ab- \left( e_{{1}}+e_{{2}} \right) ^{2}}{
4-15\, \left( e_{{1}}+e_{{2}} \right) ^{2}}}+3\,{\frac { \left( e_{{1}
}+e_{{2}} \right) ^{2}}{4-15\, \left( e_{{1}}+e_{{2}} \right) ^{2}}}.
\end{align}
Here $e_{1,2}$ are the dimensionless $\Omega$-background
\begin{align} 
e_1=-\frac{1}{e_2}=e.
\end{align}
We also scale the Coulomb moduli in order that $a$ and $b$ become dimensionless:
\begin{align} 
a_1=\hbar\,a,\, \,a_2=\hbar\,b.
\end{align}

First we compute the level-2 Shapovalov matrix of the $\mathcal{W}_3$-symmetry.
The labels for this block matrix must satisfy $|Y_L|+|Y_W|=2$.
There are therefore five choices of such a pair of Young diagrams:
\begin{align}
(Y_L,Y_W)=([2],\phi),\, (\left[1^2\right],\phi),\, ([1],[1]),\, (\phi,[2]),\, (\phi,\left[1^2\right]).
\end{align}
The level 2 matrix with these indices is given by \cite{Mironov:2009by}
\begin{align}
&\,K=\nonumber\\
&\small{\left(
 \begin {array}{ccccc} 
 4\,\Delta+\frac{c}{2}&6\,\Delta&9\,w&6\,w&{
\frac {45}{2}}\, D \Delta\\ \noalign{\medskip}6\,\Delta
&4\,\Delta\, \left( 2\,\Delta+1 \right) &6\,w \left( 2\,\Delta+1
 \right) &12\,w&27\, D \Delta+18\,{w}^{2}
\\ 
\noalign{\medskip}9\,w&6\,w \left( 2\,\Delta+1 \right) &9\, 
D {\Delta}^{2}+9\, D \Delta+9\,{w}^{2}&18\,
 D \Delta&{\frac {27}{2}}\, D w \left( 2
\,\Delta+3 \right) 
\\ \noalign{\medskip}6\,w&12\,w&18\,  D \Delta&9\,\Delta\, \left( D+1 \right) &{\frac {27}{2}}\,w
 \left( 3\,D+1 \right)
  \\ \noalign{\medskip}{\frac {45}{2}}\,  D \Delta&27\, D \Delta+18\,{w}^{2}&{\frac {27}{2
}}\, 
D w \left( 2\,\Delta+3 \right) &{\frac {27}{2}}\,w
 \left( 3\,D+1 \right) &{\frac {81}{4}}\,{D}^{2}  \Delta\,
 \left( 2\,\Delta+1 \right) +648\,{\frac {D \Delta\,
 \left( \Delta+1 \right) +4\,{w}^{2}}{22+5\,c}}
\end {array}
 \right)}.\nonumber
\end{align}
An important point is that we can factorize the level-2 Kac determinant as follows \cite{Mironov:2009by}:
\begin{align}
\label{kaclevel2}
\det{K}=
\frac
{2^43^8 \prod_{i<j}({a_{ij}}^2-e^2)^2({a_{ij}}^2-{(e+e_1)}^2)({a_{ij}}^2-{(e+e_2)}^2) }
{( {4-15e^2})^4 }.
\end{align}
Here $a_{ij}$ denotes
\begin{align}
a_{12}=a-b,\, \,a_{23}=a+2b, \,\, a_{13}=2a+b.
\end{align}

Next we calculate the $(\phi,\left[1^2\right])$-$(\phi,\left[1^2\right])$ component of the inverse matrix $K^{-1}$.
For the purpose we study the following cofactor matrix:
\begin{align}
\tilde{K}
=\left( \begin {array}{cccc} 4\,\Delta+1/2\,c&6\,\Delta&9\,w&6\,w
\\ \noalign{\medskip}6\,\Delta&4\,\Delta\, \left( 2\,\Delta+1 \right) 
&6\,w \left( 2\,\Delta+1 \right) &12\,w\\ \noalign{\medskip}9\,w&6\,w
 \left( 2\,\Delta+1 \right) &9\, D {\Delta}^{2}+9\,
 D\Delta+9\,{w}^{2}&18\,D \Delta
\\ \noalign{\medskip}6\,w&12\,w&18\,D\Delta&9\,\Delta
\, \left( D+1 \right) \end {array} \right).
\end{align}
The determinant of the matrix is
\begin{align}
&\det \tilde{K}\nonumber\\
&=-324\,c{\Delta}^{3}{w}^{2}  D +1458\,c{\Delta}^{2}{w}^{2
}  D  +486\,c\Delta\,{w}^{2}  D  +2592\,{D}^
{2}{\Delta}^{6}+2592\, D {\Delta}^{6}-2592\,{\Delta}^{4
}{w}^{2}\nonumber\\
&-9072\,\Delta\,{w}^{4}-648\,c{w}^{4}+2592\,{\Delta}^{2}{w}^{4}
+4860\,{D}^{2}{\Delta}^{4}-1620\, D{\Delta}^{4}+1620\,
{\Delta}^{2}{w}^{2}-9396\,{D}^{2}{\Delta}^{5}\nonumber\\
&+972\, D{
\Delta}^{5}-972\,{\Delta}^{3}{w}^{2}+18468\,{\Delta}^{3}{w}^{2}
 D -10044\,{\Delta}^{2}{w}^{2} D -5184\,
{\Delta}^{4}{w}^{2} D +324\,c{D}^{2}{\Delta}^{5}\nonumber\\
&+324\,c
D{\Delta}^{5}-810\,c{D}^{2}{\Delta}^{4}+486\,c
D {\Delta}^{4}-324\,c{\Delta}^{3}{w}^{2}-486\,c{D}^{2}
{\Delta}^{3}
+162\,c D{\Delta}^{3}-486\,c{\Delta}^{2}{w}^{2}\nonumber\\
&
-162\,c\Delta\,{w}^{2}+5184\,{w}^{4}.
\end{align}
By substituting (B1-4), we find the following representation:
\begin{align}
\label{cofactor}
\det \tilde{K}
=&-\frac{3^4}{{e_1}^{10}}\left( \frac{4}{4-15e^2} \right)^2\nonumber\\
&\times(128-1444{e_1}^{2}-448{e_1}^{2}ab-448{e_1}^{2}a^2-448{e_1}^{2}b^2+\cdots )
\prod_{i<j}({a_{ij}}^2-e^2).
\end{align}
 Here $(128-1444{e_1}^{2}-\cdots)$ is a polynomial of $e_1$, $a$ and $b$, which is composed of about 150 terms.
 
Finally we study the 2-instanton Nekrasov partition function.
Let us recall the 2-instanton  Nekrasov partition function for $\mathcal{N}=2$ $SU(3)$ pure Yang-Mills theory (\ref{Nekrasov2}):
\begin{align}
\label{su32inst}
Z_{SU(3),\,k=2}(a,b,c,e_1,e_2)=-\frac{9(128-1444{e_1}^{2}-448{e_1}^{2}ab-448{e_1}^{2}a^2-448{e_1}^{2}b^2+\cdots )}
{{e_1}^{10} \prod_{i<j}({a_{ij}}^2-e^2)({a_{ij}}^2-{(e+e_1)}^2)({a_{ij}}^2-{(e+e_2)}^2) }.
\end{align}
A remarkable feature is that the polynomial factors $(128-1444{e_1}^{2}-\cdots)$ of (\ref{cofactor}) and (\ref{su32inst})
are completely equal.
Hence the following identity holds for the Nakrasov partition function and the Shapovalov matrix:
\begin{align}
\label{instshap}
Z_{SU(3),\,k=2}(a,b,c,e_1,e_2)=
\frac{( {4-15e^2})^2 \det \tilde{K}}
{2^43^2 \prod_{i<j}({a_{ij}}^2-e^2)^2({a_{ij}}^2-{(e+e_1)}^2)({a_{ij}}^2-{(e+e_2)}^2) }.
\end{align}
The point is that the denominator of the right hand side is precisely that of the Kac determinant (\ref{kaclevel2}).
Then, we can prove that the our proposal also holds for 2-instanton by using (\ref{kaclevel2}) and by using (\ref{instshap}):
\begin{align}
Z_{SU(3),\,k=2}(a,b,c,e_1,e_2)&=
{\left( \frac{27}{4-15e^2}\right)}^2
\frac{ \det \tilde{K}}{ \det{K}}\nonumber\\
&={\left( \frac{-27}{4-15e^2}\right)}^2
Q_{\Delta}^{-1}(\phi, \left[1^2\right];\phi, \left[1^2\right]).
\end{align}


\end{document}